\documentclass[aps,prb,twocolumn,groupedaddress]{revtex4-1}

\usepackage{amsmath}
\usepackage{amsfonts,amssymb, graphicx}
\usepackage{wrapfig}
\usepackage[usenames,dvipsnames,svgnames,table]{xcolor}
\usepackage[colorlinks, breaklinks=true,linkcolor=red, citecolor=blue, linktocpage=true]{hyperref}

\begin{document}

\title{Dynamical Localization of Coupled Relativistic Kicked Rotors}
\author{Efim B. Rozenbaum}
\email[]{efimroz@umd.edu}
\affiliation{Joint Quantum Institute and Condensed Matter Theory Center, Department of Physics, University of Maryland, College Park, MD 20742, USA}

\author{Victor Galitski}
\affiliation{Joint Quantum Institute and Condensed Matter Theory Center, Department of Physics, University of Maryland, College Park, MD 20742, USA}

\date{\today}
\begin{abstract}
A periodically driven rotor is a prototypical model that exhibits a transition to chaos in the classical regime and dynamical localization (related to Anderson localization) in the quantum  regime. In a recent work [Phys. Rev. B {\bf 94}, 085120 (2016)], A.~C.~Keser {\it et al.} considered a many-body generalization of coupled quantum kicked rotors, and showed that in the special integrable linear case, dynamical localization survives interactions. By analogy with many-body localization, the phenomenon was dubbed dynamical many-body localization. In the present work, we study nonintegrable models of single and coupled quantum relativistic kicked rotors (QRKRs) that bridge the gap between the conventional quadratic rotors and the integrable linear models. For a single QRKR, we supplement the recent analysis of the angular-momentum-space dynamics with a study of the spin dynamics. Our analysis of two and three coupled QRKRs along with the proved localization in the many-body linear model indicate that dynamical localization exists in few-body systems. Moreover, the relation between QRKR and linear rotor models implies that dynamical many-body localization can exist in generic, nonintegrable many-body systems. And localization can generally result from a complicated interplay between Anderson mechanism and limiting integrability, since the many-body linear model is a high-angular-momentum limit of many-body QRKRs. We also analyze the dynamics of two coupled QRKRs in the highly unusual superballistic regime and find that the resonance conditions are relaxed due to interactions. Finally, we propose experimental realizations of the QRKR model in cold atoms in optical lattices.
\end{abstract}

\maketitle

\section{Introduction}
Since the discovery of  Anderson localization in 1958 \cite{Anderson58}, significant efforts---both analytical and numerical---have been made to understand how localization is affected by interactions. In 2005, Basko, Aleiner, and Altshuler~\cite{Basko06} demonstrated that, under certain conditions, localization can survive in the presence of interactions. This phenomenon was called many-body localization (MBL). The MBL state is a peculiar state of matter characterized by a number of counter-intuitive properties including ergodicity breaking \cite{Sims12}, and it has been attracting a lot of attention recently (see, e.g., Refs.~[\onlinecite{Gornyi05, *Znidaric08, *Pal10, *Ponte15, *Lazarides15}] and references therein; for reviews, see Refs.~[\onlinecite{Nandkishore15, *Altman15}]).

A different but closely related phenomenon to Anderson localization is dynamical localization. It was first introduced by Casati, Chirikov, Ford, and Izrailev \cite{Casati79, Izrailev80, Chirikov81} for a prototypical dynamical model of quantum kicked rotor (QKR)---a quantum analog of the classical kicked rotor (KR) also known as the Chirikov standard map \cite{Casati79, Chirikov79}. Experimentally, it was first observed by Moore {\it et al.} \cite{Moore95}. Dynamical localization manifests itself in quantum suppression of the chaotic classical diffusion, which for KR occurs in the angular-momentum space when the kicking strength exceeds a critical value. As opposed to Anderson localization in disordered systems, dynamical localization is not related to genuine disorder or intrinsic randomness  and is a consequence of deterministic system dynamics. However, in 1982 Fishman {\it et al.} \cite{Fishman82} showed that the QKR model can be directly mapped onto the Anderson model with quasidisorder, and that dynamical localization in QKR corresponds to  localization in the Anderson-type lattice model. In particular, in the Floquet formalism, the free rotor evolution between the kicks generates a lattice of angular-momentum states (dimensionless angular momentum is integer due to quantization on a ring), and kicking embodies hopping between the ``sites'' of this lattice.

The role of interactions in both Anderson and dynamical localization has been  studied for a long time. During a few decades, it was believed that interactions generally destroy localization due to the associated dephasing. In particular, interactions were studied directly \cite{Adachi88, *Takahashi89, Yukawa94, *Yukawa94_2, Sakagami94, *Kubotani95, Weinmann95, Shepelyansky99_2, *Shepelyansky99_3, *Shepelyansky00, Lages01, Shima04, WenLei09, *WenLei10, Boness10}, and modeled by introducing noise \cite{Shepelyansky83, Ott84, Cohen91, *Cohen91_2, *Cohen91_3, Cohen94, Shiokawa95, Borgonovi96_2, Schomerus08}, dissipation \cite{Dittrich90, *Dittrich90_2, Cohen94, Dyrting95}, and nonlinearity \cite{Benvenuto91, Shepelyansky93, *Pikovsky08, *Ermann14, Mieck04, Garcia-Mata09, *Flach09, *Mulansky09, Gligoric11, Brambila13}. Some experimental probes \cite{Graham96, Klappauf98, dArcy01, Zhang04, Duffy04, Gadway13} also tentatively suggested delocalization. However, in some special cases of two interacting QKRs, dynamical localization was found to be preserved, although weakened---specifically, for a single 2D QKR \cite{Doron88} and for the interaction potential local in rotor angular-momentum space in 1D \cite{Shepelyanskiy94, *Imry95, *Frahm95, Borgonovi95, Borgonovi96, *Shepelyansky96}. 

Although dynamical localization for two coupled QKRs was predicted to disappear in the presence of coordinate-dependent interactions \cite{Adachi88, *Takahashi89, WenLei09, *WenLei10}, in 2007 Toloui {\it et al.} \cite{Semnani07, *Semnani09} reported localized regimes in two coupled QKRs. Furthermore, recently Keser {\it et al.} \cite{Keser16} showed that coupled many-body systems can possess dynamical localization, and the corresponding phenomenon was dubbed dynamical many-body localization (DMBL). However, DMBL has been found only for a specific integrable system of linear quantum kicked rotors (LQKRs) so far, and the existence of this phenomenon in more general, nonintegrable cases remains unclear. In this paper, we propose a nonintegrable model of coupled quantum relativistic kicked rotors (QRKRs). We explicitly show dynamical localization for up to three coupled rotors (see Sec.~\ref{sec:results}), and independently of these explicit calculations we argue that the DMBL state in a many-body ensemble of such systems is possible for a wide range of parameters without fine-tuning to integrability. The many-body LQKR Hamiltonian has the form
\begin{equation}
    H^{^{\rm MB}}_{_{\rm LQKR}} = \sum\limits_{\ell=1}^L H^{^{\rm LQKR}}_{\ell}(p_{\ell}, x_\ell) + V_{\rm int}(x_1,\ldots, x_L;\, t),
\end{equation}
where the single-particle part $H^{^{\rm LQKR}}_{\ell}(p_{\ell}, x_\ell)$ defined in Eqs.~(\ref{GeneralHamiltonian}), (\ref{single_kicking}), and (\ref{Single_LQKR}) depends on angular momentum linearly as $C_\ell p_\ell$. And this many-body Hamiltonian is a high-angular-momentum limit of the many-body QRKR Hamiltonian given by
\begin{equation}
    H^{^{\rm MB}}_{_{\rm QRKR}} = \sum\limits_{\ell=1}^L H^{^{\rm QRKR}}_{\ell}(p_{\ell}, x_\ell) + V_{\rm int}(x_1,\ldots, x_L;\, t),
\end{equation}
where $H^{^{\rm QRKR}}_{\ell}(p_{\ell}, x_\ell)$ defined in Eqs.~(\ref{GeneralHamiltonian}), (\ref{QRKR_freeHamiltonian}), and (\ref{single_kicking}) depends on angular momentum as $\sqrt{(C_\ell p_\ell)^2 + M_{\ell}^2}$, since
\begin{equation}
	\sqrt{(Cp)^2 + M^2} \simeq Cp, \;\;\text{as}\;\; Cp/M \rightarrow \infty
\end{equation}
(parameter definitions are given below). Therefore, any possible delocalization stops at sufficiently high angular momenta where this asymptotic dominates. On the other hand, outside of this asymptotic regime, the classical counterpart of the QRKR model exhibits chaotic behavior, and a quantum Anderson-type mechanism is necessary to induce localization. We show that this mechanism also works to some extent in the presence of interactions and conclude that in the general many-body case, localization can result from an interplay of both effects.

Apart from the dynamical localization, we also address regimes where the single relativistic kicked rotor exhibits novel transport behavior and examine them in the interacting case. We find that interactions facilitate this behavior and increase the number of such regimes.

New interesting transport effects can be found if the asymptotic behavior of the dispersion relation at low angular momentum is different from the behavior at high angular momentum. In 2003, Matrasulov {\it et al.}  \cite{Matrasulov03, *Matrasulov05} suggested a model of QRKR, a quantum version of a classical relativistic kicked rotor (RKR) \cite{Chernikov89}. Both RKR and QRKR models naturally possess this dispersion property. It is important to note that from the viewpoint of the dispersion relation, QRKR interpolates between conventional QKR at low angular momenta and exactly solvable LQKR \cite{Fishman82_2, Prange84, Berry84, Simon85} at high angular momenta. Recently, Zhao {\it et al.} \cite{Zhao14} discovered rich transport properties of RKR and QRKR that included various regimes ranging from localization  to superballistic transport.

In general, the transport properties can be classified by the value of the index $\nu$ in the time dependence of the mean-squared generalized ``coordinate.'' For a rotor, the relevant choice is the angular-momentum space:
\begin{equation}\label{Eq:transport}
\langle p^2 \rangle \sim t^\nu.
\end{equation}
In the case of pure localization $\nu = 0$. The values of $\nu > 0$ correspond to various types of delocalization. $\nu = 1$ corresponds to the standard diffusion $\langle p^2 \rangle \sim t$, while the case $\nu \neq 1$ is called anomalous diffusion. In particular, $\nu \in (0, 1)$ is dubbed subdiffusion and $\nu \in (1, 2)$ is superdiffusion. The regime with $\nu = 2$ corresponds to ballistic transport. 
There is also a special, less studied case of transient anomalous diffusion called superballistic transport that corresponds to $\nu > 2$. Only a few examples of this regime are known to date \cite{Hufnagel01, Zhang12, Stutzer13, Iubini15}. Interestingly, both RKR and QRKR exhibit the superballistic transport regime \cite{Zhao14}.

Besides the angular-momentum dynamics, QRKR also naturally possesses a spin-like degree of freedom, and dynamics in this ``spin'' space is quite peculiar (strictly speaking, it is the particle-antiparticle space of the 1D Dirac equation, but we will refer to it as spin for brevity). The first spinful kicked rotor model---spin-$1/2$ QKR---was suggested by Scharf~\cite{Scharf89} and later studied in Refs.~[\onlinecite{Tahaha94, *Ossipov04, *Dahlhaus11, *Chen14, *Tian16}], but the evolution of the spin in either of models did not receive much attention.

In the present paper, first we review the QRKR model and introduce its spin dynamics properties that, to the best of our knowledge, have not been discussed in the literature. Then we numerically demonstrate robust localization upon driving for the model with up to $3$ interacting QRKRs. Most importantly, it means that in this model interparticle coupling that corresponds to infinite-range interaction in the respective lattice model does not always destroy few-body localization (as opposed to the case in Refs.~[\onlinecite{Adachi88, *Takahashi89, Yukawa94, *Yukawa94_2, Sakagami94, *Kubotani95, Weinmann95, Shepelyansky99_2, *Shepelyansky99_3, *Shepelyansky00, Lages01, Shima04, WenLei09, *WenLei10, Boness10, Shepelyansky83, Ott84, Cohen91, *Cohen91_2, *Cohen91_3, Cohen94, Shiokawa95, Borgonovi96_2, Schomerus08, Dittrich90, *Dittrich90_2, Cohen94, Dyrting95, Benvenuto91, Shepelyansky93, *Pikovsky08, *Ermann14, Mieck04, Garcia-Mata09, *Flach09, *Mulansky09, Gligoric11, Brambila13, Klappauf98, dArcy01, Gadway13}], but similarly to the one in Refs.~[\onlinecite{Semnani07, *Semnani09}]). More generally, we show that the coupled model inherits the transport regimes of the single QRKR model. If generalized to a many-body system of QRKRs, this statement results in DMBL similar to that  found in Ref.~[\onlinecite{Keser16}], but for a nonintegrable system. Independently of the numerics, but in agreement with it, we argue that this is the case because the difference between the dynamics of the many-body QRKRs model and the integrable model considered in Ref.~[\onlinecite{Keser16}] vanishes as the angular-momentum terms increase and overwhelm the mass terms.

\section{Quantum Relativistic Kicked Rotor \label{sec:QRKR}}
In this section, we review the QRKR model (see also Ref.~[\onlinecite{Zhao14}]) and study the spin dynamics and spin-momentum entanglement in this model. We find a number of unusual dynamic regimes involving the spin. From this point on, we mostly refer to the rotors' angular momenta simply as momenta for shortness.

As for any kicked system, the Hamiltonian of the QRKR model reads
\begin{equation} \label{GeneralHamiltonian}
\hat{H}(t) = \hat{H}_0 + V\sum\limits_{n=-\infty}^\infty\delta(t-n),
\end{equation}
where $t$ is a dimensionless time (measured in the units of the kicking period, $T$). Throughout the paper, we use the notation 
\begin{equation}
\Delta(t) \equiv \sum\limits_{n=-\infty}^\infty \delta(t-n).
\end{equation}
For QRKR, the free part $\hat{H}_0$ is the dimensionless 1D Dirac Hamiltonian:
\begin{equation} \label{QRKR_freeHamiltonian}
\hat{H}_0(p) = 2\pi\alpha p\hat{\sigma}^x + M\hat{\sigma}^z,
\end{equation}
where $2\pi\alpha \equiv C$ plays the role of an effective speed of light, $M$ is an effective mass, and $\boldsymbol{\hat\sigma}$ is a vector of Pauli matrices. $p$ is a dimensionless angular momentum operator, $p = -\dot{\imath}\frac{\partial}{\partial x}$, and $x$ is an angular coordinate of the particle, $x \in [0, 2\pi)$. We assume that kicking has the following form:
\begin{equation} \label{single_kicking}
V(x) = K\cos(qx),
\end{equation}
where $K$ is an effective kicking strength, and $q \in \mathbb{N}$ specifies the spatial period of the potential. Note that the kicking potential (\ref{single_kicking}) is proportional to the unit matrix in the spin space. 
\begin{figure}[t]
\includegraphics[width=\linewidth]{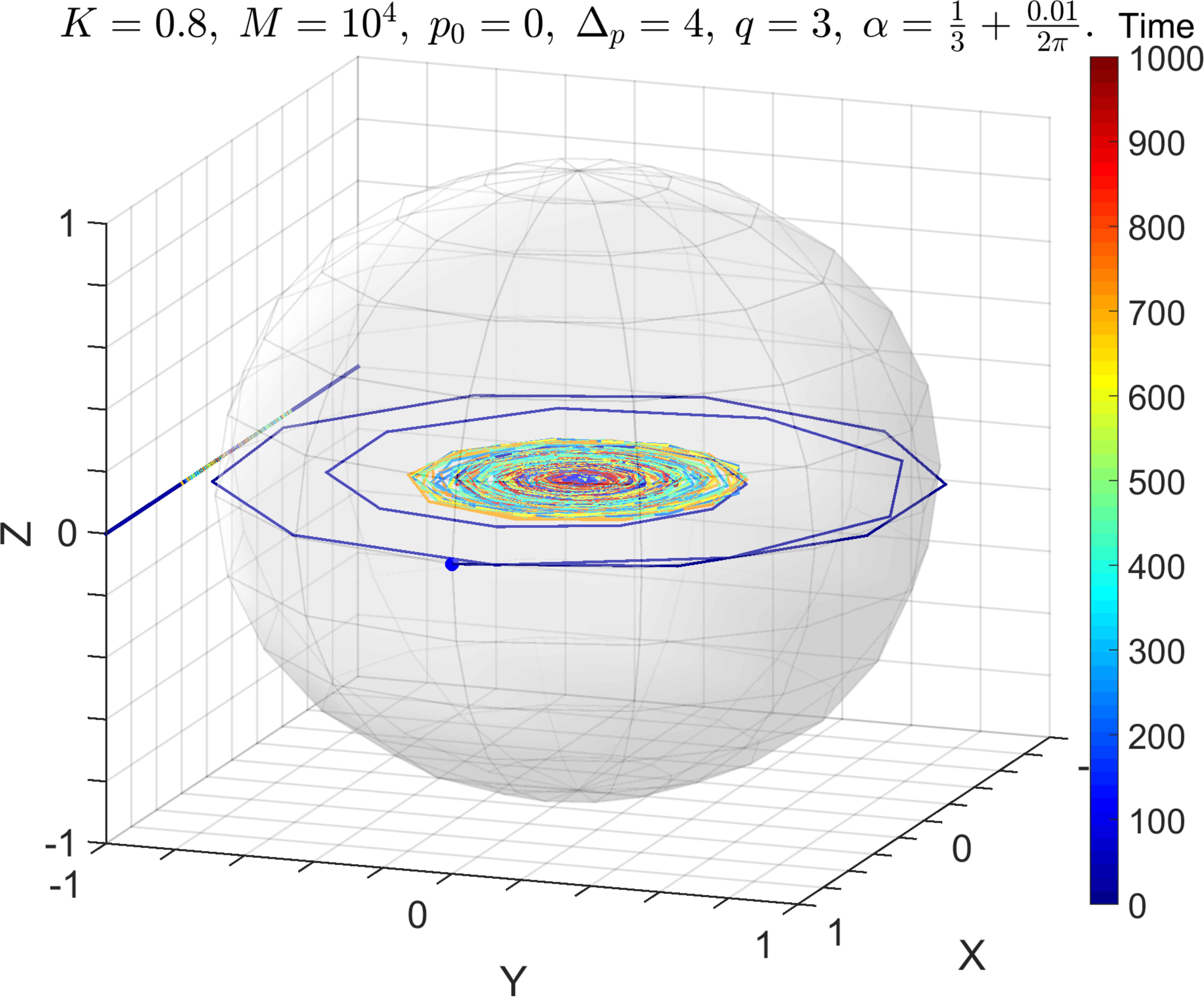}
\caption{\label{Large_M_q_3} Spin dynamics with a large mass, zero average initial momentum, and small initial momentum-distribution width $\Delta_p$ (parameters are shown above the figure). The blue marker indicates the initial point, and the color indicates time (in the units of the kicking period). The projection of the spin trajectory is shown on the $XZ$ coordinate plane.}
\end{figure}
\begin{figure}[t]
\includegraphics[width=\linewidth]{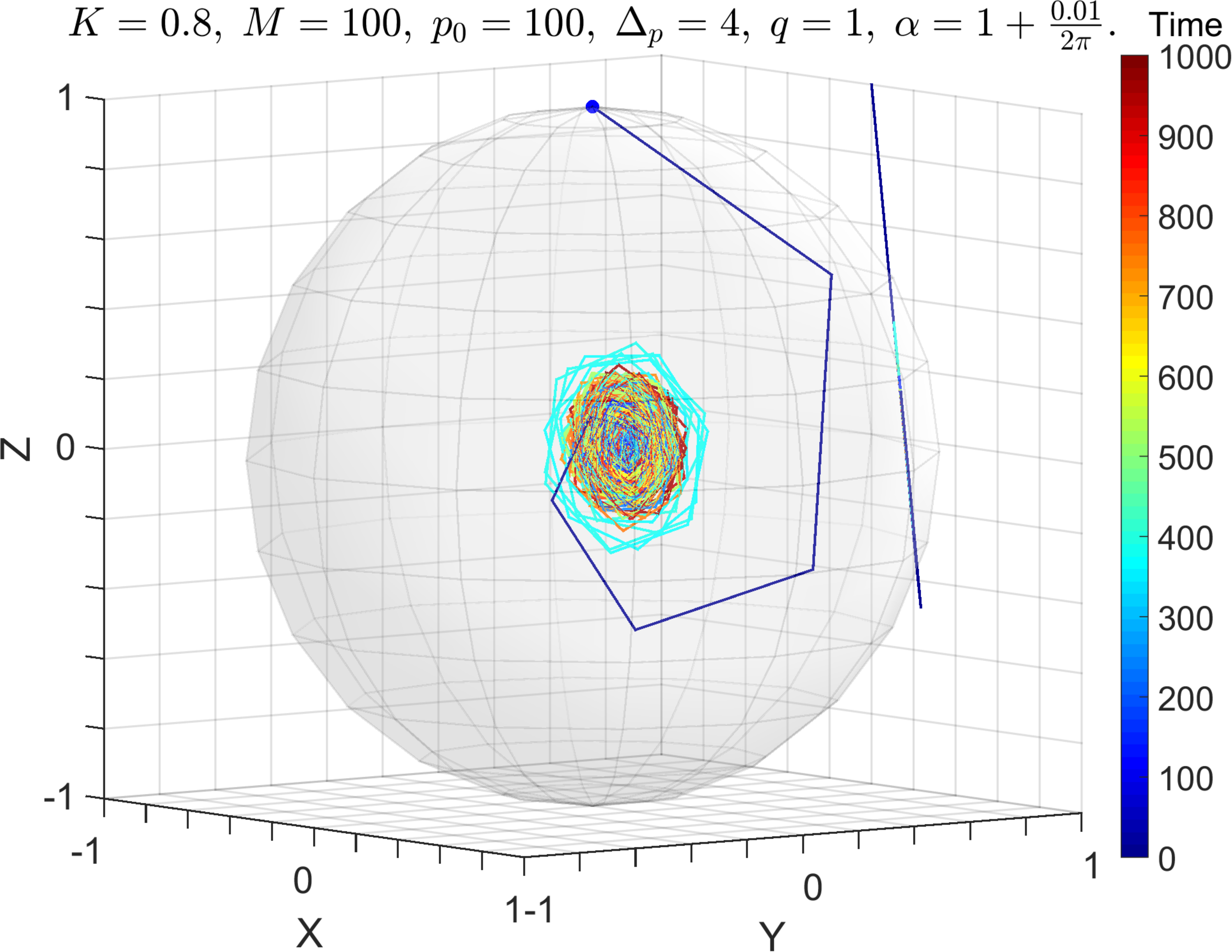}
\caption{\label{Medium_M_Med_p0} Spin dynamics with a medium mass equal to the average initial momentum, and small initial momentum-distribution width $\Delta_p$ (parameters are shown above the figure). The blue marker indicates the initial point, and the color indicates time. The projection of the spin trajectory is shown on the $XZ$ coordinate plane.}
\end{figure}

In order to quantify the role of quantum and relativistic effects in QRKR as compared to RKR and QKR, respectively (see Sec.~\ref{sec:results} for the coupled rotors case), we make connection to the actual Dirac equation for a kicked relativistic spin-$1/2$ particle of mass $m$ confined to a 1D ring of radius $R$:
\begin{equation}
\hspace{-4pt}
i\hbar\frac{\partial}{\partial t_p}\Psi = \left[ cp_p\hat{\sigma}^x + mc^2\hat{\sigma}^z + kR\cos(qx)\Delta\left(\frac{t_p}{T}\right)\right]\Psi,\hspace{-3pt}
\end{equation}
where $t_p = tT$ is physical time, $c$ is the speed of light, $p_p = \hbar p/R$ is physical (linear) momentum, and $k$ is the amplitude of the kicking force.
Introduce a dimensionless ``effective Planck constant'':
\begin{equation}
\hbar_{\rm eff} = \frac{\hbar T}{mR^2}.
\end{equation}
In the dimensionless Dirac equation, we absorb $\hbar_{\rm eff}$ into the other parameters, so that the Hamiltonian (\ref{GeneralHamiltonian}), (\ref{QRKR_freeHamiltonian}), (\ref{single_kicking}) enters it as follows:
\begin{equation} \label{Dirac_Equation}
\dot{\imath}\frac{\partial}{\partial t} \Psi = \left[Cp\hat{\sigma}^x + M\hat{\sigma}^z + K\cos(qx)\Delta(t)\right]\Psi,
\end{equation}
where
\begin{eqnarray} \label{dimless_alpha}
	C &\equiv& 2\pi\alpha = \dfrac{cT}{R}, \\\label{dimless_M}
	M &=& \dfrac{mc^2T}{\hbar} = \dfrac{C^2}{\hbar_{\rm eff}}, \\
\label{dimless_K}	K &=& \dfrac{kTR}{\hbar} = \dfrac{K_{_{\rm RKR}}}{\hbar_{\rm eff}},
\end{eqnarray}
and $K_{_{\rm RKR}}$ is the dimensionless kicking strength in the nonquantum RKR model [see Eq.~(\ref{RKR_Ham}) below].
Note that this straightforward interpretation is not related to the feasible physical realizations. Some of the latter are proposed in Sec.~\ref{sec:proposals}.

Consider integer times $t$ only. Since the Hamiltonian is periodic in time---$H(t+1) = H(t)$---and the external potential has a kicking form, the stroboscopic single-period evolution of the wave functions governed by Eq.~(\ref{Dirac_Equation}) is given by the Floquet operator $\hat{F}$ as
\begin{equation} \label{evol}
\Psi(t+1) = \hat{F}\Psi(t),
\end{equation}
where
\begin{align} \label{Floquet}
\hat{F} &= \exp\left[-i\hat{H}_0(p)\right]\exp\left[-iV(x)\right] \\ \nonumber
&= \exp\left[-i\left(2\pi\alpha p\hat\sigma^x + M\hat\sigma^z\right)\right]\exp\left[-iK\cos(qx)\right].
\end{align}

\begin{figure}[t]
\includegraphics[width=\linewidth]{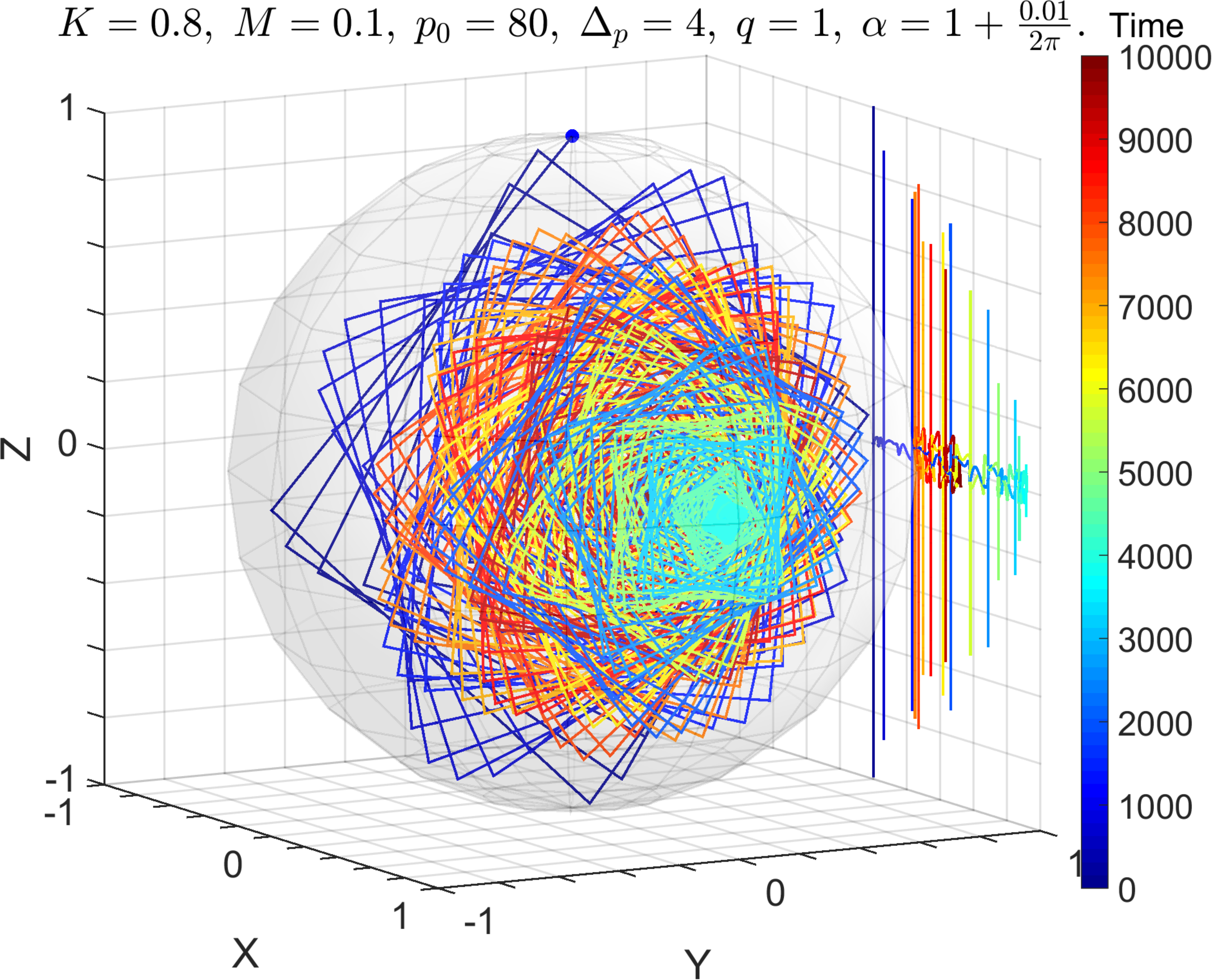}
\caption{\label{Small_M_High_p0} Spin dynamics with a small mass, high average initial momentum, and medium initial momentum-distribution width $\Delta_p$ (parameters are shown above the figure). The blue marker indicates the initial point, and the color indicates time. The projection of the spin trajectory is shown on the $XZ$ coordinate plane.}
\end{figure}
\begin{figure}
\includegraphics[width=\linewidth]{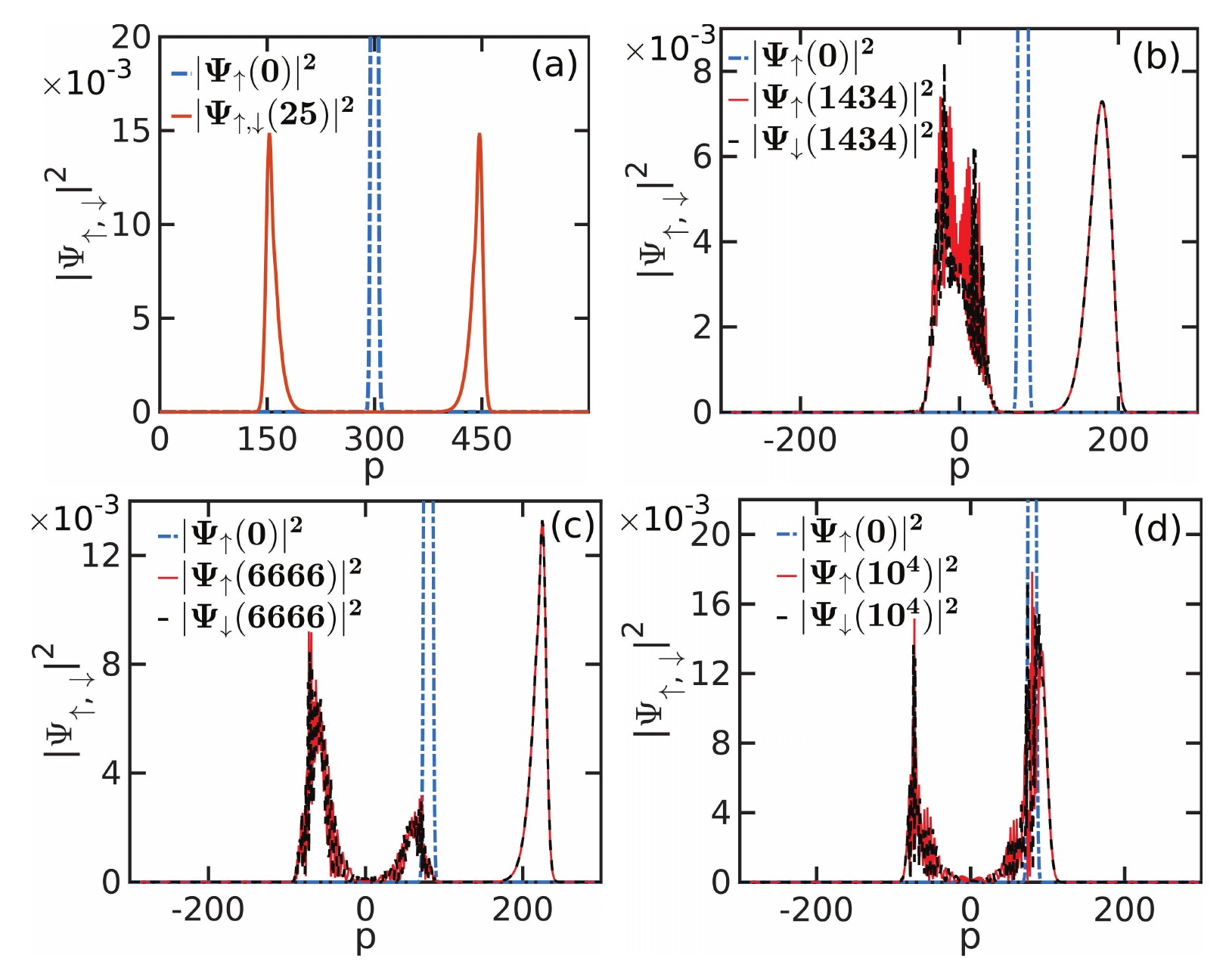}
\caption{\label{Small_M_High_p0_WF} Wave function at different stages of evolution with the same parameters as in Fig.~\ref{Small_M_High_p0}. Panel (a) shows the initial wave function (dot-dashed blue line) and that at $t=25$ (solid red line) in case of starting at $p_0 = 300$---very far from $p = M/(2\pi\alpha)$. Panels (b)--(d) exactly correspond to Fig.~\ref{Small_M_High_p0} and show the initial wave function (dot-dashed blue lines) and the up and down spinor components (solid red and dashed black lines, respectively) at the times indicated in the parentheses. The initial down component is zero.}
\vspace{20pt}
\includegraphics[width=\linewidth]{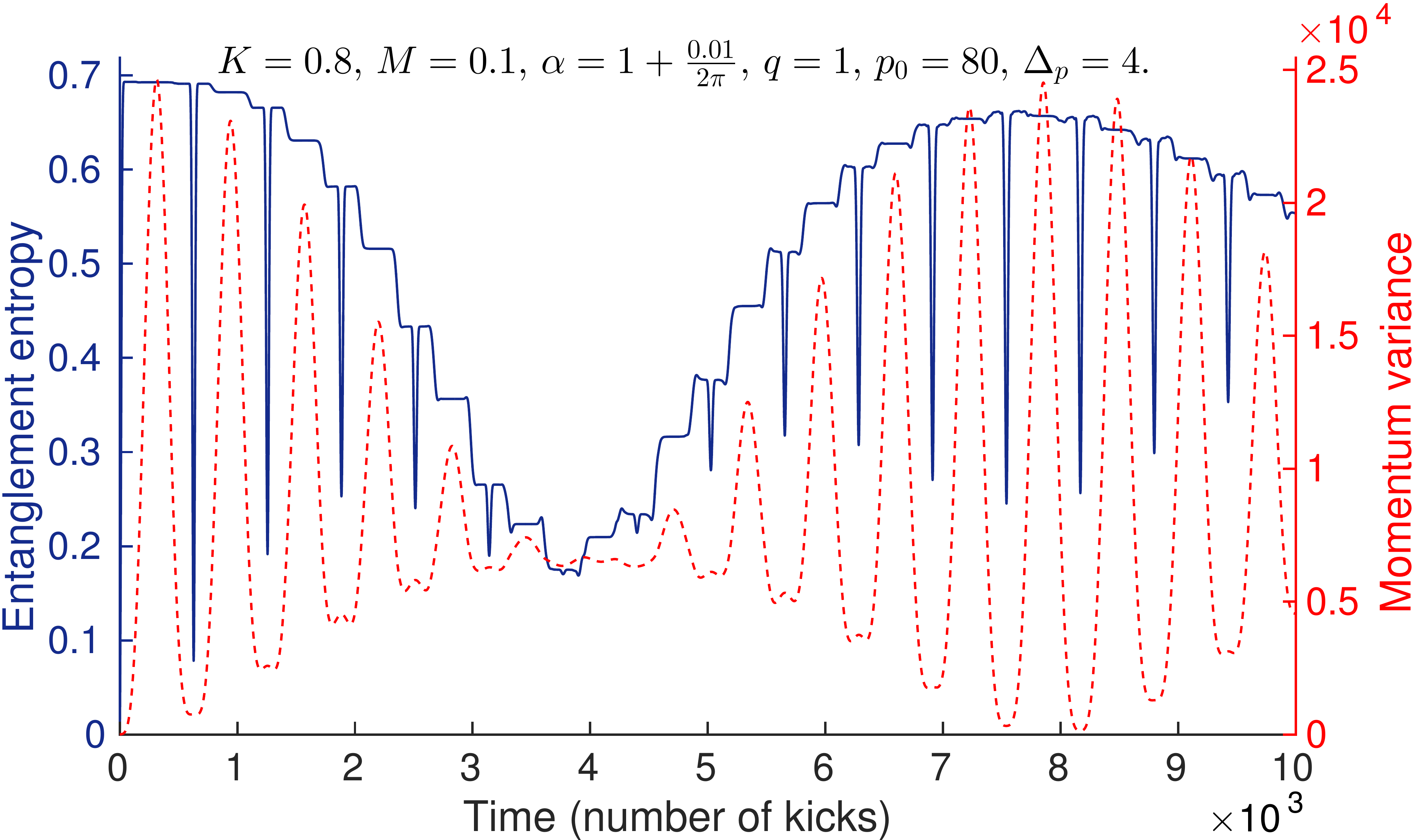}
\caption{\label{EntEnt} Entanglement entropy (solid blue line, left axis) and momentum variance $(\langle p^2\rangle - \langle p \rangle^2)$ (dashed red line, right axis) in the process corresponding to Fig.~\ref{Small_M_High_p0}. Parameters are shown above the figure.}
\end{figure}

An efficient way of calculating the evolution (\ref{evol}) numerically is by switching from the coordinate representation to the momentum representation back and forth, applying each part of the Floquet operator (\ref{Floquet}) in its eigenbasis. This approach allowed us to reproduce the results for single QRKR obtained in Ref.~[\onlinecite{Zhao14}]. The details of the numerical implementation of this method are given for coupled QRKRs in Sec.~\ref{TwoQRKRs}.

As we mentioned above, in the high-momentum region the QRKR model can be approximated by the LQKR model. It is determined by  Hamiltonian  (\ref{GeneralHamiltonian}) with the free part:
\begin{equation} \label{Single_LQKR}
H_0^{\mathrm{LQKR}}(p) = 2\pi\alpha p
\end{equation}
and kicking potential (\ref{single_kicking}). This model has been proved to be integrable in any dimension by Figotin and Pastur~\cite{Figotin84}.

During the free evolution between the kicks, the local eigenspinors of $\hat{H}_0(p)$, i.e. the eigenspinors at any given $p$ as a parameter, acquire the phases $\varphi_{\pm}(p)~=~\pm~\sqrt{(2\pi\alpha p)^2 + M^2}$, where the quantized momentum $p$ takes only integer values. As discussed in Ref.~[\onlinecite{Zhao14}], the transport regime in QRKR is determined by the phases $\varphi_{\pm}(p)$ along with the kicking potential parameter $q$ and initial conditions. Specifically, in the low-momentum region ($2\pi\alpha p \ll M$), QRKR is always localized (for the same reason as QKR and LQKR), because propagators $\exp[-\dot{\imath}\varphi_{\pm}(p)]$ act as quasi-random-number generators for a sequence of integers $p$. In the high-momentum limit, QRKR tends to LQKR, and the behavior of $\exp\left[-\dot{\imath}\varphi_{\pm}(p)\right]$ is determined by the rationality of $\alpha$. Namely, irrational values of $\alpha$ give rise to the localized phase, and rational values---$\alpha = r/s \; (r, s \in \mathbb{Z} \text{ are relatively prime})$---lead to delocalization (in particular, to ballistic transport) if $q/s \in \mathbb{Z}$. In the remaining case of $\alpha = r/s$, but $q/s \notin \mathbb{Z}$, the dynamics is bounded, but this is not related to the Anderson-type localization \cite{Berry84, Zhao14}. In the general case of the wave function containing components with $2\pi\alpha p \sim M$, an additional pattern---the superballistic transport---arises due to the leakage of the wave function from the low-momentum region to the high-momentum one \cite{Zhao14}. 
Specifically, following the qualitative argument from Ref.~[\onlinecite{Zhao14}], there are three contributions to the momentum-space transport. One contribution is constant and comes from localization in the disordered region at small momenta. Another one is ballistic; i.e., the momentum variance grows quadratically in time. It comes from the periodic nondisordered (in terms of $\exp[-\dot{\imath}\varphi_{\pm}]$) high-momentum region. And the third contribution is superballistic. It is related to the transfer of population from the moderate-momentum to the high-momentum region and can be qualitatively described by $\int\limits_0^tdt'\Gamma(t-t') Dt'^2$, where $\Gamma(t)$ is a population transfer rate and $D$ is a coefficient of ballistic transport. In the simplest case of $\Gamma(t) \approx {\rm const}$, this integral readily gives cubic growth of momentum variance.

\begin{figure}
\includegraphics[width=\linewidth]{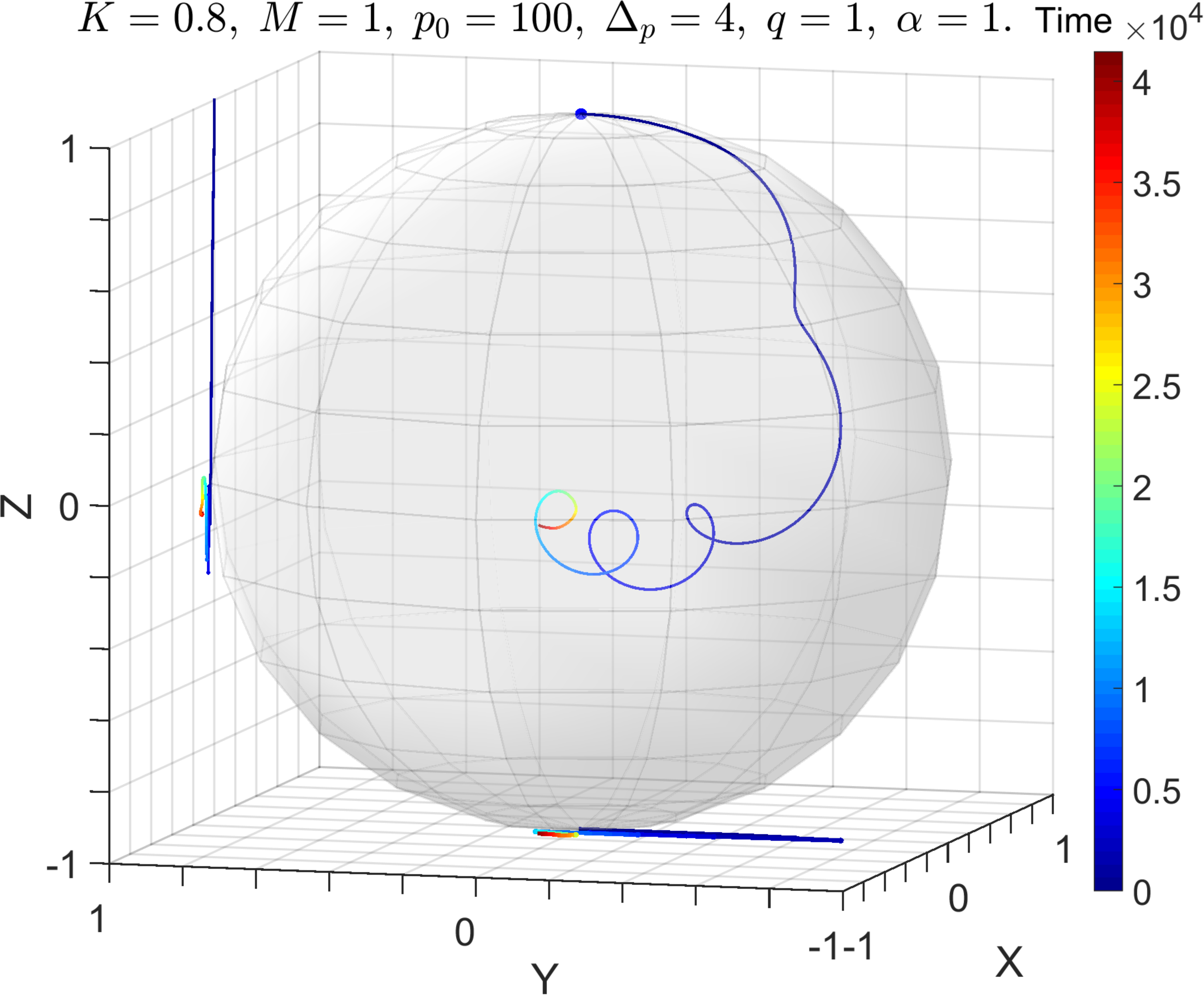}
\caption{\label{ResonanceSpin} Spin dynamics in a resonant regime corresponding to delocalization ($\alpha$ is rational). Parameters are shown above the figure. The blue marker indicates the initial point, and the color indicates time. The projections of the spin trajectory are shown on the $XY$ and $XZ$ coordinate planes.}
\end{figure}

Besides the rich dynamics that QRKR shows in the momentum space, it also possesses very peculiar patterns in spin dynamics, even if the kicking is spin-independent, as in Eq.~(\ref{single_kicking}). These patterns are related to the entanglement between the spin and momentum degrees of freedom that occurs at each step as a result of the combination of free evolution and kicking. We performed a series of calculations of the spin evolution in the case of spin-independent kicking (\ref{single_kicking}). In Figs.~\ref{Large_M_q_3}~--~\ref{Small_M_High_p0} and \ref{ResonanceSpin}, we show trajectories of the tip of the spin vector---more precisely, the vector $ 2\,\mathbf{s}(t) = \left< \Psi(t)\left|\boldsymbol{\hat\sigma} \right| \Psi(t) \right> $---within the Bloch sphere in four representative regimes. Figs.~\ref{Large_M_q_3}~--~\ref{Small_M_High_p0} correspond to the localized phase, while Fig.~\ref{ResonanceSpin} describes the spin dynamics in the delocalized phase. As an initial state, we chose a Gaussian $\Psi(p, t=0) \sim \exp\left[-(p-p_0)^2/(2\Delta_p^2)\right]\chi_\sigma$, where $\chi_\sigma = \left|\uparrow\right> + i \left|\downarrow\right>$ in Fig.~\ref{Large_M_q_3} and $\chi_\sigma = \left|\uparrow\right>$ in all other cases.

In Fig.~\ref{Large_M_q_3}, $M/2\pi\alpha$ is much larger than the initial momentum spread centered around zero, and due to localization, the mass remains two orders of magnitude greater than $2\pi\alpha p$ for the highest populated momentum components. In this case, the spin-tip trajectory is a flat disk that lies in the $XY$ plane and constitutes rotation via $\exp\left[-\dot{\imath}M\hat\sigma^z\right]$. The radius of the spin trajectory is determined by the degree of the spin-momentum entanglement and is oscillating in time within certain boundaries.

In Fig.~\ref{Medium_M_Med_p0}, the mass $M$ and the initial momentum $p_0$ are equal, which leads to the flat trajectory tilted at an angle---$\tan(\theta) \approx M/(2\pi\alpha p_0) = 1/(2\pi\alpha)$---to the $YZ$ plane. When the ratio between $2\pi\alpha p_0$ and $M$ is varied, the trajectory remains flat in a certain range of parameters, and only tilt angle changes accordingly.

In Fig.~\ref{Small_M_High_p0}, $2\pi\alpha p_0 \gg M$, and the momentum spread $\Delta_p \gg M$. In this case, we have alternating regimes of dynamics. When the  majority of the momentum-space population is far away from $p=M/(2\pi\alpha) \ll p_0$, the spin-tip trajectory is flat---in this case, it constitutes rotation in the $YZ$ plane due to the action of $\exp\left[-\dot{\imath}2\pi\alpha p\hat\sigma^x\right]$. The wave function dynamics far away from $p=M/(2\pi\alpha)$ consists of periodic splitting into two parts and recombining back. One of these parts corresponds to classical acceleration due to in-phase kicking and another one---to deceleration due to out-of-phase kicking. As a manifestation of localization, these parts span only very limited vicinity of the initial wave packet; Fig.~\ref{Small_M_High_p0_WF}(a) shows these parts at the largest separation alongside the initial state. When the wave function components are split, the spin tip stays very close to some point $X_0$ at the $X$ axis. And when the components recombine, the spin tip comes to the surface of the Bloch sphere developing a flat part of the trajectory that is parallel to the $YZ$ plane and crosses the point $X_0$. However, if the initial wave function is centered close enough to $p=M/(2\pi\alpha)$, as is the case in Fig.~\ref{Small_M_High_p0}, one of the split components goes through this point, and the spin tip starts to move along the $X$ axis until that component leaves the vicinity of $p = M/(2\pi\alpha)$. Once the components of the wave function recombine, a new flat disk parallel to $YZ$ plane develops, and then this periodic pattern continues with the $X$-motion until the next disk is generated. When the motion along the $X$ axis brings the spin tip to the surface of the Bloch sphere, the $X$-motion reflects off it and continues back to the center of the Bloch sphere. During this motion, each time a nonvanishing part of the wave function passes through $p = M/(2\pi\alpha)$, it gets modulated, split, and eventually becomes very noisy in that region [see Figs.~\ref{Small_M_High_p0_WF}(b)--\ref{Small_M_High_p0_WF}(d)]. In Fig.~\ref{EntEnt}, we show the entanglement entropy as a function of time corresponding to this case. It is defined as
\begin{equation}
  S(t) = -{\rm tr}\left[\rho_s(t)\ln\rho_s(t)\right],\vspace{-10pt}
\end{equation}
where\vspace{-15pt}
\begin{equation}
  \rho_s(t) = \sum\limits_{p=-\infty}^\infty\left|\Psi(p,t)\right>\left<\Psi(p,t)\right|.
\end{equation}
\begin{figure}[t]
\includegraphics[width=\linewidth]{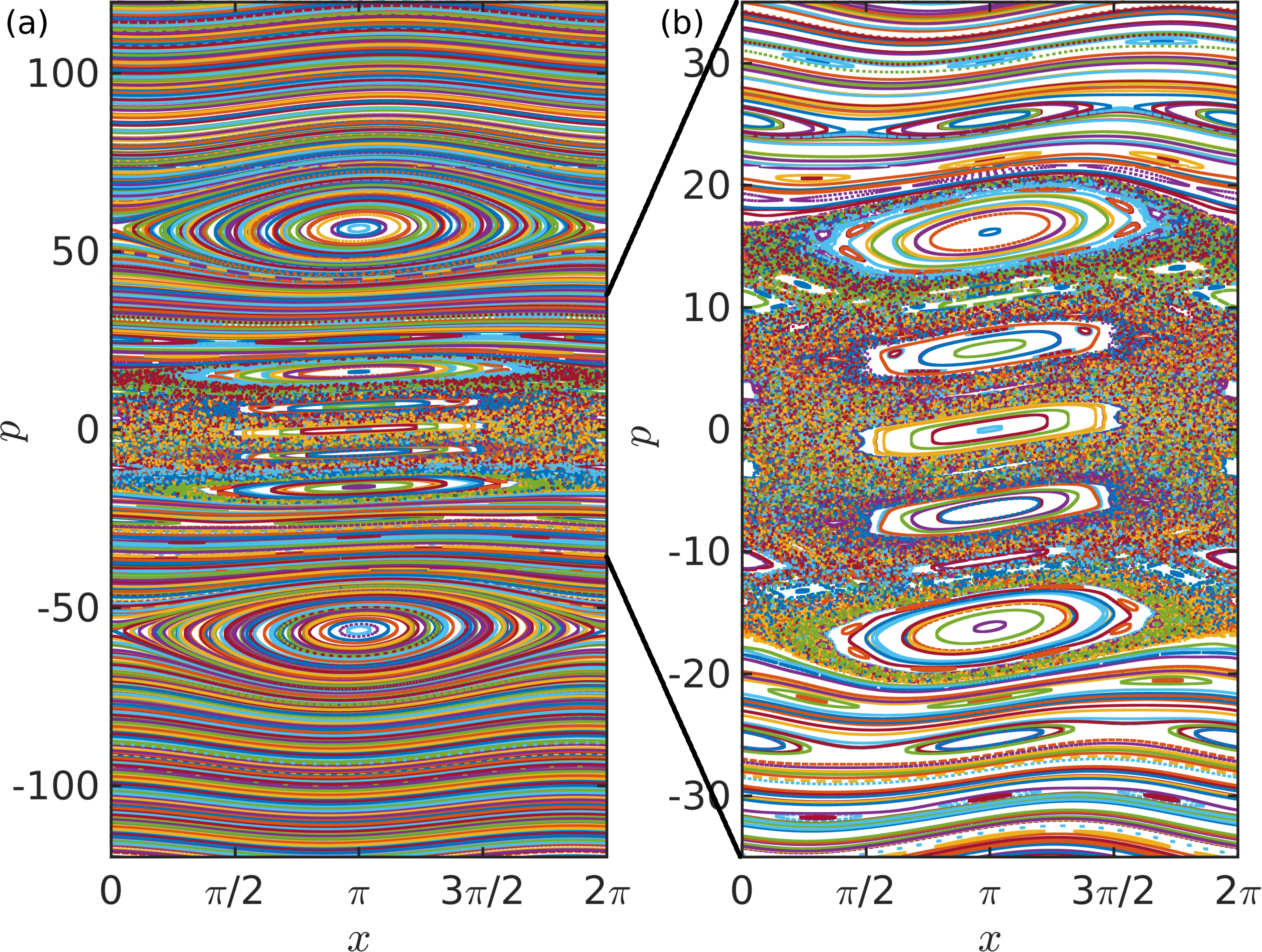}
\caption{\label{PhaseMapWeak} Phase map of RKR in a moderate kicking regime. Parameters: $C=20$, $K_{_{\rm RKR}}=2$, $q=1$. Panel (b) shows zoomed region of moderate angular momenta from panel (a). At low momenta, there are both chaotic and periodic trajectories. At high momenta, all trajectories are periodic.}
\vspace{20pt}
\includegraphics[width=\linewidth]{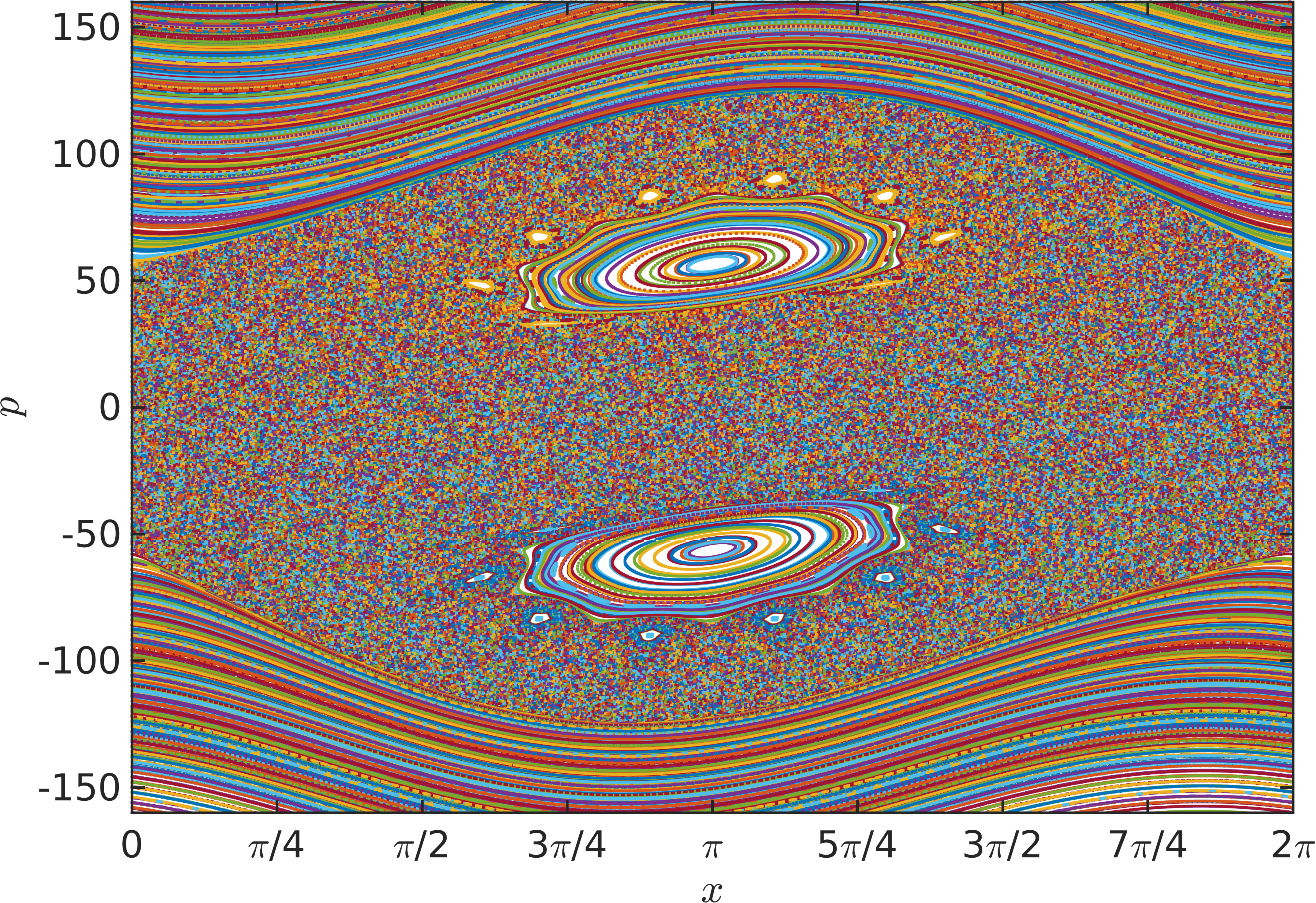}
\caption{\label{PhaseMapStrong} Phase map of RKR in a strong kicking regime. Parameters: $C=20$, $K_{_{\rm RKR}}=20$, $q=1$. At low momenta, chaotic trajectories span most of the phase space. At high momenta, all trajectories are periodic.}
\end{figure}
The sharply pronounced dips in the entanglement entropy correspond to the flat disk structures in the Bloch sphere with their edges coming close to the surface of the sphere. And the envelope of the entanglement entropy corresponds to the motion of the spin tip along the $X$ axis. As the time goes, the wave function becomes more and more noisy, and no sharp disk structures are generated for some time. This corresponds to the region between $3000$ and $5000$ kicks in Figs.~\ref{Small_M_High_p0} and \ref{EntEnt}. However, at some point, the revival of the disks structure in the Bloch sphere occurs, and the corresponding revival of the dips structure in the entanglement entropy can be seen. As the wave function goes through the point $p = M/(2\pi\alpha)$ many times, it splits again, and becomes randomized, so that eventually, the motion within the Bloch sphere becomes less regular. However, it retains the features described above for at least as long as $2\times 10^4$ kicks.

Figure~\ref{ResonanceSpin} shows the spin dynamics in the case of resonant value of $\alpha$, i.e., delocalization in the momentum space. In this regime, the motion of the spin tip is continuously slowing down. As time goes, more and more momentum components get populated, and the spin-tip trajectory tends to one limiting point inside the Bloch sphere.

\begin{figure}[t]
\includegraphics[width=\linewidth]{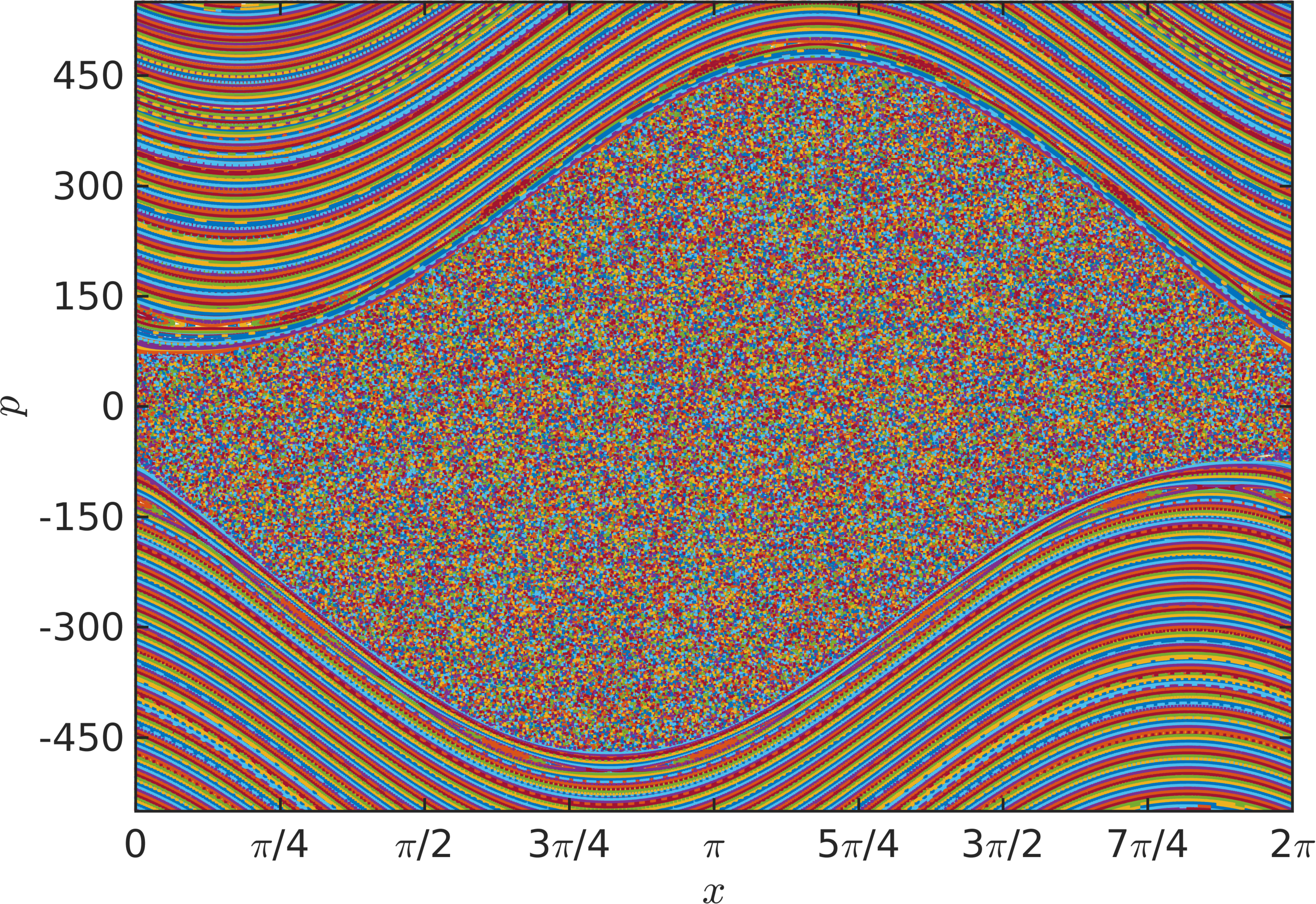}
\caption{\label{PhaseMapVeryStrong} Phase map of RKR in a very strong kicking regime. Parameters: $C=20$, $K_{_{\rm RKR}}=200$, $q=1$. At low momenta, all trajectories are chaotic. At high momenta, all trajectories are still periodic.}
\end{figure}
\section{Critical kicking strength in RKR}
Before discussing coupled QRKRs, we need to address the difference between the classical models of RKR and nonrelativistic KR with respect to the notion of critical kicking strength. RKR is described by the dimensionless Hamiltonian:
\begin{equation} \label{RKR_Ham}
H_{_{\rm RKR}} = C^2\sqrt{1 + \frac{p^2}{C^2}} + K_{_{\rm RKR}} \cos(qx)\Delta(t).
\end{equation}
According to Ref.~[\onlinecite{Chernikov89}], as opposed to KR, in RKR, different Kolmogorov-Arnold-Moser (KAM) tori in the phase space are destroyed at different critical values of the kicking strength $K^{i, \rm cr}_{_{\rm RKR}}$, which depend on the parameter $C$. Most importantly, there are global limiting tori at high momentum that do not get destroyed at any finite value of $K_{_{\rm RKR}}$ if $C/2\pi \equiv \alpha \notin \mathbb{Z}$ (the latter condition is always satisfied in the quantum localized phase described in Sec.~\ref{sec:QRKR}, i.e., when $\alpha \notin \mathbb{Q}$). This behavior is illustrated in Figs.~\ref{PhaseMapWeak}~--~\ref{PhaseMapVeryStrong}. Variegated regions of moderate momentum filled with chaotic trajectories are always bounded from both sides by global regular trajectories that span the rest of the phase space.

However, although the existence of the limiting tori guarantees classically bounded trajectories, it is not the only source of localization exhibited by the QRKR, even when coupling is introduced. The QRKR shows localization within both classically regular and classically chaotic regions. Therefore, in general, localization is caused by a combination of the classically bounded phase space and quantum Anderson-type localization. The same argument holds for coupled QRKRs. We illustrate it in Sec.~\ref{sec:results}. 

\section{Two and Three Coupled Quantum Relativistic Kicked Rotors \label{TwoQRKRs}}
The many-body generalization of the LQKR model was considered in Ref.~[\onlinecite{Keser16}], and it was shown analytically that the many-body LQKR model may exhibit the DMBL phase. In other words, it was shown that localization may survive in the presence of interactions. This finding partially motivated the present study of the few-body generalization of the QRKR model, which is qualitatively distinct from the LQKR model due to the nonintegrability.

In this section, we consider the simplest interacting cases: a two-body and a three-body coupled-QRKR models. Specifically, we consider the models with the Hamiltonian
\begin{equation}
\hat{H}(t) = \hat{H}_0 + (V + H_{\rm int})\Delta(t),
\end{equation}
where $\hat{H}_0$ is chosen in two different ways. First, for the two-body case, we consider the sum of the Dirac Hamiltonians of the noninteracting QRKRs:
\begin{equation}\label{Eq:spinful_H0}
\hat{H}_0 = 2\pi\alpha_1 p_1 \hat\sigma^x_1 + M_1 \hat\sigma^z_1 + 2\pi\alpha_2  p_2\hat\sigma^x_2 + M_2 \hat\sigma^z_2.
\end{equation}
Here $\hat\sigma^{x,z}_1 = \hat\sigma^{x,z} \otimes \mathbb{I}$ \;and\; $\hat\sigma^{x,z}_2 = \mathbb{I}\otimes \hat\sigma^{x,z}$ are $4\times4$ matrices, $C_\ell \equiv 2\pi\alpha_\ell$. Second, we use a spinless version of the QRKR to construct
\begin{equation}\label{Eq:spinless_H0}
\hat{H}_0 = \sum\limits_{\ell = 1}^L\sqrt{(2\pi\alpha_\ell p_\ell)^2 + M_\ell^2}, \;\;\; L=2,\; 3.
\end{equation}
As mentioned in Sec.~\ref{sec:QRKR}, it has been shown in Ref.~[\onlinecite{Zhao14}] that the spinless model $H'_{_{\rm QRKR}} = \sqrt{(Cp)^2 + M^2} + K\cos(qx)\Delta(t)$ possesses the same localization properties as the spinful QRKR. For three coupled particles, we only use the spinless $\hat{H}_0$ [Eq.~(\ref{Eq:spinless_H0})] to reduce computational complexity.
The kicking potential has a standard form:
\begin{equation}
V(x_1,\ldots, x_L) = \sum\limits_{\ell=1}^L K_\ell\cos(qx_\ell), \;\;\; L=2,\; 3.
\end{equation}
The interaction part is chosen in the same way as for coupled QKRs in Refs.~[\onlinecite{Semnani07, *Semnani09}], which generalizes the potentials considered in Refs.~[\onlinecite{Adachi88, Doron88}]:
\begin{eqnarray}\nonumber
H_{{\rm int}} = \dfrac{1}{2}\sum\limits_{j,\ell=1}^L K^{\rm int}_{j\ell}\{\cos(qx_j)\cos(qx_\ell)& + \cos[q(x_j - x_\ell)]\},&\\
&L=2,\;3.&
\end{eqnarray}
We study these models numerically, and show that the localized phase is persistent with respect to the interaction in a certain range of parameters.

The Floquet operator
\begin{equation} \label{Floquet-two}
\hat{F} = \exp\left[-i\hat{H}_0\right]\exp\left[-i(V + H_{\rm int})\right].
\end{equation}
For $\hat{H}_0$ in Eq.~(\ref{Eq:spinful_H0}), $\Psi$ is a four-component function. Its four components correspond to the four possible spin configurations of two particles: $\uparrow\uparrow, \; \uparrow\downarrow,\; \downarrow\uparrow,\; \text{and} \downarrow\downarrow$, respectively. The free part of the Floquet operator (\ref{Floquet-two}) in the spinful case is calculated using the properties of the Pauli matrices:
\begin{eqnarray} \label{mom_floquet} \nonumber
&&\exp\left[-i\hat{H}_0\right] = \bigotimes\limits_{\ell=1}^2\left[\cos\left(\sqrt{C_\ell^2p^2_\ell + M_\ell^2}\right)\mathbb{I}\right. \\
&&\left.-i\sin\left(\sqrt{C_\ell^2p^2_\ell + M_\ell^2}\right)\dfrac{C_\ell p_\ell \hat\sigma^x + M_\ell \hat\sigma^z}{\sqrt{C_\ell^2p^2_\ell + M_\ell^2}} \right]
\end{eqnarray}
and can be efficiently applied numerically to a four-component wave function on a momentum-space grid at each step. Similarly, the kicking and interaction part of the evolution operator (\ref{Floquet-two}) can be efficiently applied numerically to a wave function on a coordinate-space grid.

We choose an initial wave function in the basis of momentum eigenstates:
\begin{equation}
	\Psi(t=0) = \sum\limits_{P}a_P(0) \left|P\right>,
\end{equation} 
where $P = \{p_1, p_2\}$ or $\{p_1, p_2, p_3\}$
and, to make one step in time, we perform the discrete Fourier transform of $\{a_P\}$ to go to the coordinate representation, where the potential part of the Floquet operator (\ref{Floquet-two}) is diagonal, and we apply this part to the vector representing the coordinate-space wave function. After that, we perform the inverse Fourier transform to go back to the momentum representation and apply the free part [operator (\ref{mom_floquet}) for the spinful case] to it. Then this cycle starts over for the next step. This scheme allows us to achieve efficient numerical evolution that only requires application of diagonal operators and a fast Fourier transform.

\section{Numerical experiments with two and three coupled QRKR\lowercase{s} \label{sec:results}}
\begin{figure}
\includegraphics[width=\linewidth]{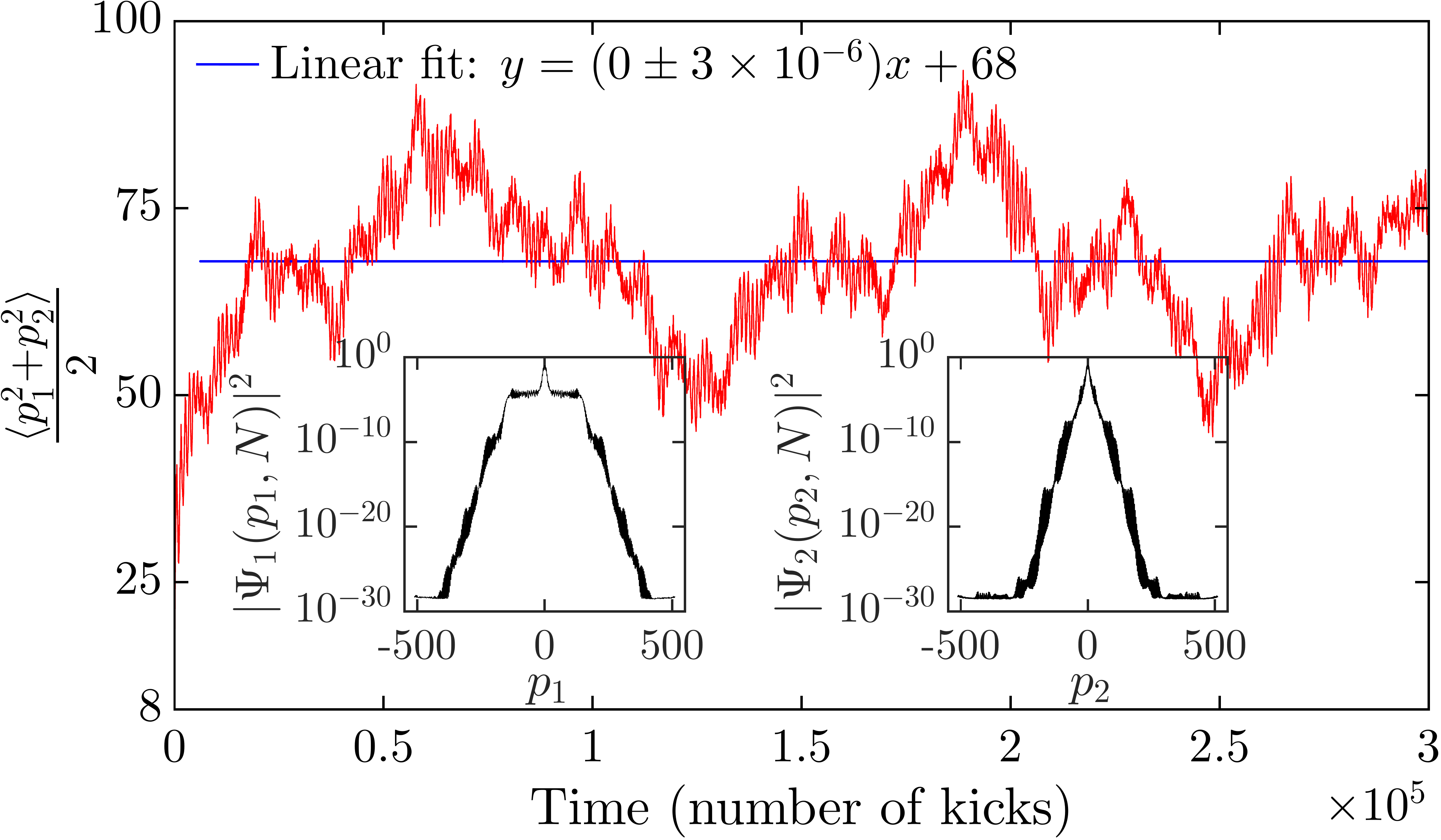}
\caption{\label{LocalizationStandard} Average mean-squared momentum per particle in the localized regime. Oscillating red line shows the calculation result. Straight blue line is a linear fit that shows no slope up to a fitting error (equation is given inside the plot). Parameters: $\alpha_1~=~1/3~+~0.02/2\pi,\,\alpha_2~=~1/3~-~0.03/2\pi,\, q~=~3,\\ K_1~=~K_2~=~0.8,\, K^{\rm int}_{12}~=~0.04,\, M_1~=~M_2~=~12$ ($\hbar_{{\rm eff}_1}~\approx~\hbar_{{\rm eff}_2}~\approx~0.37$). Insets: Probability density at the final time $t~=~N~=~3\times10^5$ kicks as a function of each momentum while integrated over the other one.}
\end{figure}

\begin{figure}
\includegraphics[width=\linewidth]{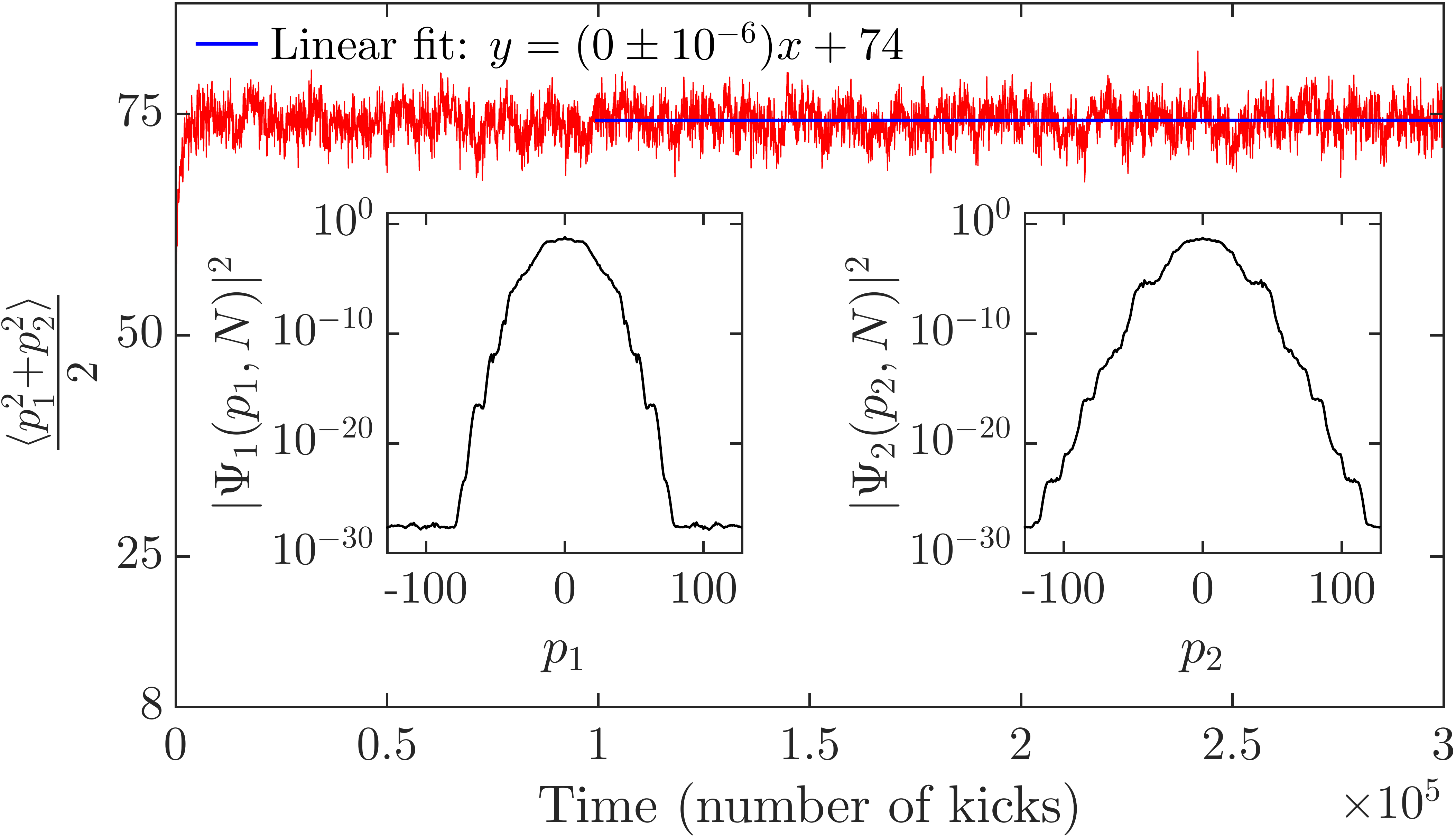}
\caption{\label{Localization_heff_1} Average mean-squared momentum per particle in the localized regime. Oscillating red line shows the calculation result. Straight blue line is a linear fit that shows no slope up to a fitting error (equation is given inside the plot). Parameters: $\hbar_{\rm eff}~=~1,\,\, \alpha_1~=~1.6~+~0.1/2\pi,\,\, \alpha_2~=~1.6~-~0.15/2\pi,\,\, q~=~1,\\ K_{1,2}~=~4/\hbar_{\rm eff},\,\, K^{\rm int}_{12}~=~0.2/\hbar_{\rm eff},\,\, M_{1,2}~=~\left(2\pi\alpha_{1,2}\right)^2/\hbar_{\rm eff}$. Insets: Probability density at the final time $t~=~N~=~3\times10^5$ kicks as a function of each momentum while integrated over the other one.}
\end{figure}

\begin{figure}[t]
\includegraphics[width=\linewidth]{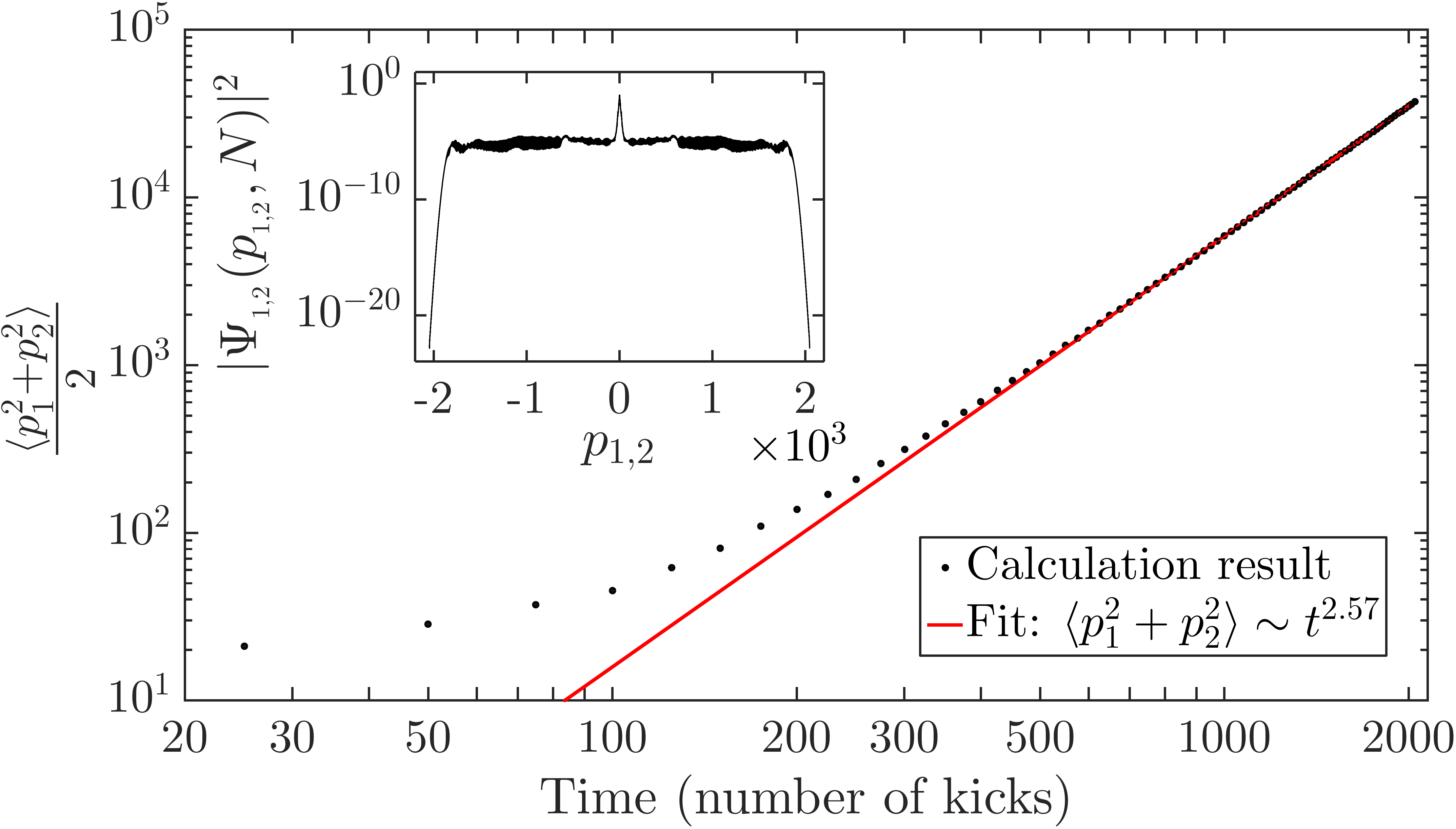}
\caption{\label{Resonance} Average mean-squared momentum per particle in the superballistic regime. Black points show the calculation result. Red line is a power fit. Parameters: $\alpha_{1,2}~=~1/3~-~0.03/2\pi,\,\, q~=~3,\,\, K_{1,2}~=~0.7,\,\, K^{\rm int}_{12}~=~0.2,\\ M_{1,2}~=~12$. Inset: Probability density at the final time $t~=~2\times10^3$ kicks as a function of each momentum while integrated over the other one.}
\end{figure}

In Figs.~\ref{LocalizationStandard}~--~\ref{Resonance}, we present time dependence of the average mean-squared momentum per particle for two coupled spinful QRKR particles: $\langle p_1^2+p_2^2\rangle/2$. In Ref.~[\onlinecite{Keser16}], this quantity was shown to be a reliable indicator of dynamical localization (as opposed to the average energy). 
Different values of the parameters determine various regimes that are exhibited by our model. We start in a Gaussian-shaped initial state in the momentum space centered around the point $\left(p_1^{(0)}, p_2^{(0)}\right) = (0, 0)$ with both spins up:
\begin{equation}
\Psi(p_1, p_2, t=0) \sim \exp\left[-\frac{p_1^2}{2\Delta_1^2} -\dfrac{p_2^2}{2\Delta_2^2}\right] \left|\uparrow\uparrow\right>,
\end{equation}
and widths $\Delta_1 = \Delta_2 = 4$. Therefore, the initial value of the average mean-squared momentum per particle is $\langle p_1^2+p_2^2\rangle/2 = 8$.
Figures~\ref{LocalizationStandard} and \ref{Localization_heff_1} represent the dynamically localized state. It is characterized by nonresonant values of the velocities ($\alpha_1\neq \alpha_2 \notin \mathbb{Q}$). Saturation of the average mean-squared momentum per particle is verified by linear fits that have zero slope up to a fitting error; the corresponding equations are shown inside the plots. The insets in these figures show the probability density at the final time $\left|\Psi(p_1, p_2, t=N)\right|^2$ as a function of each of the momenta while integrated over the other one. As one can see from these insets, in the regime of localization, the wave functions decay exponentially with momenta and, in the vicinity of numerical boundaries, reach the values below $10^{-27}$. This ensures that during the evolution, the population does not come close to the numerical boundaries, and there is no unphysical reflection from them. Stable exponential decay of a wave function with a constant bound on its width indicates localization.
In Fig.~\ref{LocalizationStandard}, we take parameters similar to those used in Ref.~[\onlinecite{Zhao14}] for the single QRKR and add $5\%$ of interaction, i.e., $K^{\rm int}_{12} = 0.05 K_{1,2}$. These parameters correspond to $\hbar_{\rm eff_1} \approx \hbar_{\rm eff_2} \approx 0.37$. In Fig.~\ref{Localization_heff_1}, we set $\hbar_{\rm eff_1} = \hbar_{\rm eff_2} = 1$ and periodicity parameter $q = 1$ and obtain more stable localization. Other parameters in this case are such that in the corresponding RKR model, many tori are destroyed giving way to a broad chaotic region. In particular, the kicking strength constant exceeds the first (and, in this case, the only) critical constant---$\hbar_{\rm eff} K_{1,2} > K^{\rm 1, cr}_{_{\rm RKR}} \approx 2$---corresponding to the single RKR. Figure \ref{Resonance} shows the delocalized phase described by the resonant values of the velocities: $\alpha_1 = \alpha_2$. Even though $\alpha_1, \alpha_2 \notin \mathbb{Q}$---so that for single QRKR, such $\alpha$ guarantees localization---according to Ref.~[\onlinecite{Keser16}], in the many-body LQKR model, equal values of the velocities correspond to additional resonances due to the interaction that lead to divergence of the emergent momenta-containing integrals of motion (IOMs) present in the many-body LQKR model~\cite{Keser16}. 
In the many-body QRKR model at large momenta, these IOMs become approximate. Nevertheless, their divergence results in the divergence of the associated momenta, which is confirmed by our numerical results. In particular, in Fig.~\ref{Resonance} we see a rapid transport in the momentum space that causes the fast growth of the probability density near the numerical boundaries of the momentum grid. Unfortunately, this complication makes further numerical analysis at large time scales inefficient. However, we can clearly see the signs of superballistic transport in this plot. In particular, in this example, $\langle p_1^2+p_2^2\rangle \sim t^{2.57}$ until the wave function reaches the grid boundaries at the time beyond $t=2000$ kicks, and we can not rely on the numerics after that point. The inset shows the probability density at final time $\left|\Psi(p_1, p_2, t=N)\right|^2$ as a function of each of the momenta while integrated over the other one ($p_1$ and $p_2$ dependencies are the same in this case due to symmetry).

\begin{figure}[t]
\includegraphics[width=\linewidth]{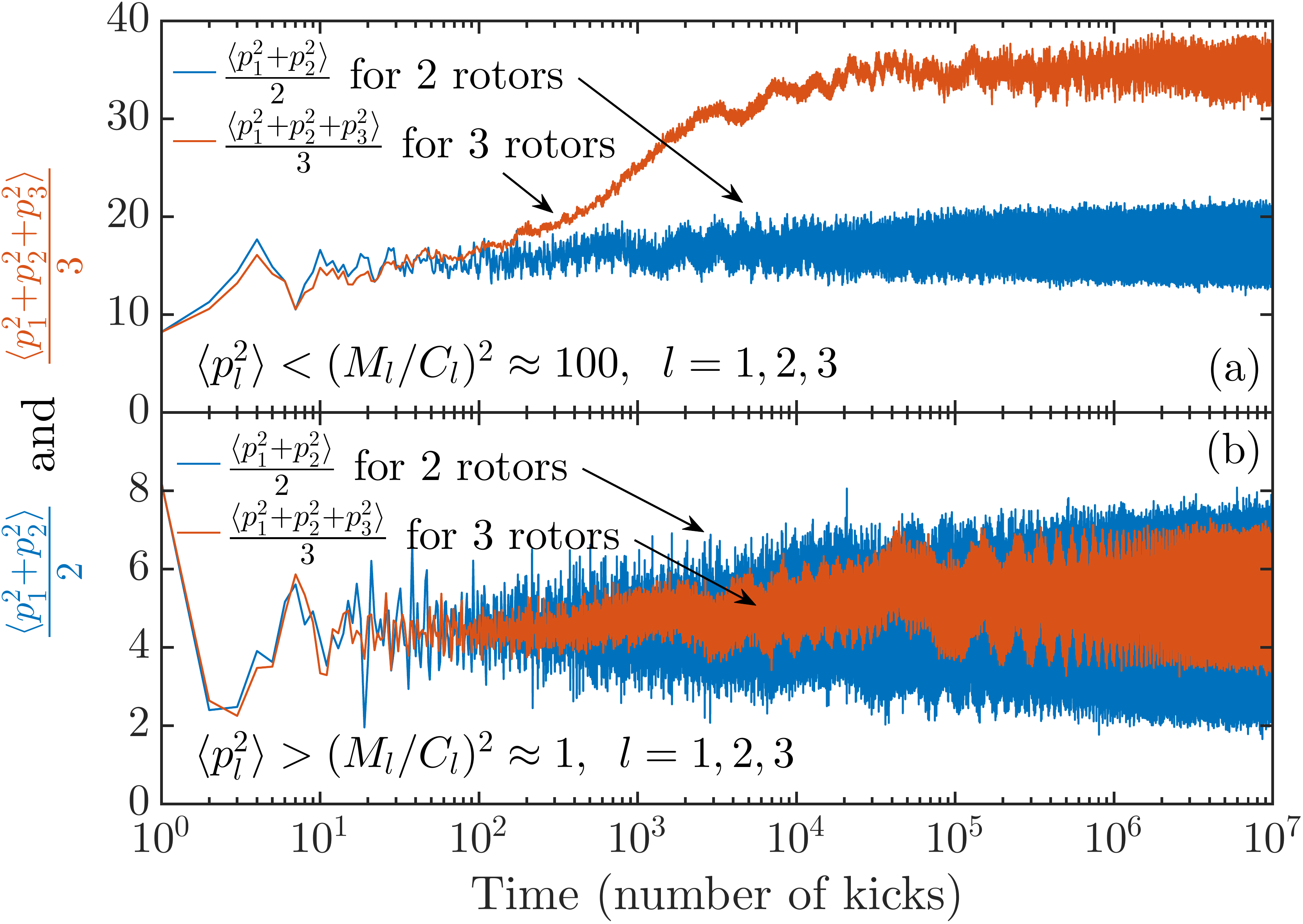}
\caption{\label{Fig:two_vs_three_spinless} Average momentum variance per particle in the localized regime for two (blue lines) and three (red lines) coupled spinless QRKRs. Panel (a) shows localization far from integrable limit; parameters are $\hbar_{\rm eff}=1,\,\, \alpha_1=1.6+0.1/2\pi,\,\, \alpha_2=1.6-0.15/2\pi,\,\,(\alpha_3=1.6+0.225/2\pi),\,\, q=1,\,\, K_1=2/\hbar_{\rm eff},\,\,K_2=3/\hbar_{\rm eff},\,\,(K_3=1.5/\hbar_{\rm eff}),\,\, K^{\rm int}_{12}=0.1/\hbar_{\rm eff},\,\, (K^{\rm int}_{23}=0.07/\hbar_{\rm eff},\,\, K^{\rm int}_{31}=0.13/\hbar_{\rm eff}),\,\, M_{1,2, (3)}=\left(2\pi\alpha_{1,2,(3)}\right)^2/\hbar_{\rm eff}$. Panel (b) shows localization close to the integrable limit. Parameters are the same as in panel (a) except for $\alpha_{1,2,(3)}$ being multiplied by a factor $0.1$.}
\end{figure}

In Fig.~\ref{Fig:two_vs_three_spinless}, we compare localization in two-particle and three-particle spinless QRKR models at the same respective parameters. The plots are given in the lin-log scale to show details of saturation. We should note that upon increasing the interaction strengths, there appear regimes of long-lasting logarithmic growth of the average mean-squared momentum per particle that may be generic for coupled nonintegrable dynamical systems but also satisfy the condition $\nu = 0$ in Eq.~(\ref{Eq:transport}). As one can see, in Fig.~\ref{Fig:two_vs_three_spinless}, panel (a), the localization length and time it takes the mean-squared momentum to saturate increase with the number of particles, as expected given the increased contribution of interactions. However, this panel shows saturation well below the integrable region determined by $p^2_\ell \gg (M_\ell/C_\ell)^2 = (C_\ell/h_{{\rm eff}_\ell})^2 \approx 100$. In contrast to it, in panel (b), the near-integrability threshold is $p^2_\ell \gg 1$, and the saturated value satisfies this condition to some extent. Notice that in this case, the saturated values of mean-squared momentum per particle are in the same range for two and three particles.

\section{Experimental Proposal \label{sec:proposals}}
Due to the structure of the Floquet operators (\ref{Floquet}) and (\ref{Floquet-two}) for single and coupled QRKRs respectively, as well as for any kicked Hamiltonian, quantum dynamics is almost invariant with respect to swapping the free and the kicked parts; i.e., the Hamiltonians
\begin{equation} \label{non_swapped}
\mathcal{H} = H_0 + H_1\Delta(t) \vspace{-5pt}
\end{equation}
\vspace{-5pt}and
\begin{equation}  \label{swapped}
\mathcal{H}_{\rm swap} = H_0\Delta(t) + H_1
\end{equation}
generate the same Floquet evolution. More precisely, in order to get completely the same dynamics, when swapping, one should also change from considering evolution between the moments of time right after the kicks to those just before the kicks and vice versa. 

In particular, the single QRKR is equivalent to a model with the Hamiltonian
\begin{equation} \label{QRKR_exp}
\hat{H}^{\rm QRKR}_{\rm swap} = \left( 2\pi\alpha p\hat\sigma^x + M\hat\sigma^z \right)\Delta(t) + K\cos(qx).
\end{equation}
Let us put $q = 1$. Recall that the angular coordinate $x \in [0, 2\pi)$ and the dimensionless angular momentum is quantized---$p \in \mathbb{Z}$. Then one can establish a correspondence between the QRKR and a spin\nobreakdashes-$\frac{1}{2}$ particle  hopping on a 1D lattice subject to a pulsed magnetic field. This correspondence is summarized in the following table.
\renewcommand{\arraystretch}{2.5}
		\begin{center}
  			\begin{tabular}{ c  c  c  c  c  c }
			\hline\hline
   	 		QRKR & $x$ & $p$ & $2\pi\alpha$ & $M$  & $K$\\ 
  			Spin-$\frac{1}{2}$ &\; $ka+\pi$ \; &\; $j$ \;&\; $-\dfrac{\mu T}{2\hbar} B_x(1)$ \;&\; $-\dfrac{\mu T}{2\hbar} B_z$ \;&\; $T\dfrac{\mathcal{T}}{\hbar}$ \\
 				 \hline\hline
 			 \end{tabular}
		\end{center}
\renewcommand{\arraystretch}{1}
Here $k$ is a quasi-momentum in the first Brillouin zone for a lattice with real-space site numbers $j$ and a lattice constant $a$. $\mu$ is a magnetic moment associated with a particle's spin, $B_x(1)$ is an $x$ component of the magnetic field on the site $j=1$ [so that in general, $B_x(j) = jB_x(1)$ is linear in real space], $B_z$ is a uniform $z$ component of the magnetic field, and $\mathcal{T}$ is a hopping energy. So, we get the following 1D single-band tight-binding Hamiltonian for a spin\nobreakdashes-$\frac{1}{2}$ particle that is being periodically kicked via the external magnetic field:
\begin{equation} \label{latt_exp}
H_{\rm magn} = -\mu\left[B_x(j) s_x + B_z s_z\right]\Delta(t) - \mathcal{T}\cos(ka),
\end{equation}
where $s_x$ and $s_z$ are the particle's spin components.

If we keep the kicking function $\Delta(t)$ at the original place---as in Eq.~(\ref{non_swapped})---we get a Hamiltonian for a spin\nobreakdashes-$\frac{1}{2}$ particle in a time-independent magnetic field and in a pulsed optical lattice:
\begin{equation} \label{lp_exp}
H_{\rm pl} = -\mu\left[B_x(j) s_x + B_z s_z\right] - \mathcal{T}\cos(ka)\Delta(t).
\end{equation}
In this case, however, it is important to keep the lattice on so as not to recover the quadratic kinetic energy term. It can be done by switching from the deep optical lattice to the shallow one back and forth instead of turning it on and off completely.

Another possible setup for implementing the QRKR model is a two-level atom in a laser field with detuning $\delta$ and nonuniform Rabi frequency $\Omega(j)$ at the $j^{\rm th}$ site in the presence of a pulsed optical lattice. It is implemented via the following mapping.
\renewcommand{\arraystretch}{2.5}
		\begin{center}
  			\begin{tabular}{ c  c  c  c  c  c }
			\hline\hline
   				 QRKR & $x$ & $p$ & $2\pi\alpha$ & $M$  & $K$\\ 
  				 Atom & \; $ka+\pi$ \;&\; $j$ \;&\; $T\Omega(1)$ \;&\; $-T\dfrac{\delta}{2}$ \;&\; $T\dfrac{\mathcal{T}}{\hbar}$ \;\\
 				 \hline\hline
 			 \end{tabular}
		\end{center}
		\renewcommand{\arraystretch}{1}
Then in the rotating wave approximation,
       \begin{align} \nonumber
       H_{\rm at} = \hbar\Omega(j) \left(|g\rangle\langle e| + |e\rangle\langle g|\right) &- \hbar\dfrac{\delta}{2} \left(|e\rangle\langle e| - |g\rangle\langle g|\right)& \\  \label{opt_exp} &- \mathcal{T}\cos(ka)\Delta(t),&
       \end{align}
where $|g\rangle$ and $|e\rangle$ are the ground and excited states of the atom in the rotating frame. The same caveat regarding the quadratic kinetic term as in the previous setting applies here. As well as in the previous examples, one could kick the first part of this Hamiltonian instead of applying a pulsed lattice.

Similar models can be constructed on the basis of the QKR and LQKR models. In particular,
\begin{equation} \label{QKR_exp}
H^{\rm QKR}_{\rm swap} = \dfrac{\hbar_{\rm eff}p^2}{2}\Delta(t) + K\cos(x)
\end{equation}
and
\begin{equation} \label{LQKR_exp}
H^{\rm LQKR}_{\rm swap} = 2\pi\alpha p \Delta(t) +  K\cos(x)
\end{equation}
correspond to a charged particle in a 1D lattice in the presence of a kicked electric field. This field is linear in space for $H^{\rm QKR}_{\rm swap}$ and uniform for $H^{\rm LQKR}_{\rm swap}$.

Extensions of the single-particle models (\ref{QRKR_exp}) -- (\ref{LQKR_exp}) to the case of many interacting particles can be mapped to corresponding many-body QRKRs, QKRs, or LQKRs in the same way.

Hamiltonians (\ref{QRKR_exp}) -- (\ref{LQKR_exp}) might be realized in cold atoms in optical lattices. Interestingly, according to the mapping $p \mapsto j$, for such systems, dynamical localization as well as other intriguing transport regimes such as superballistic transport, take place in real space rather than in momentum space, which makes these phenomena especially demonstrative in experiment.

\section{Conclusion}
Starting with the single-particle QRKR model that possesses the rich variety of transport phases, we introduced its peculiar spin dynamics phenomenology and generalized it to the model of interacting QRKRs. For the models of two and three coupled QRKRs, we showed that the transport regimes---and, in particular, the localized phase---can survive interactions. We are not aware of any previous study of coupled QRKRs, but we point out that for the well-studied coupled QKRs and related static lattice models, most works predict delocalization at least for infinite-range interaction. Our calculations indicate the existence of the localized regimes for such a coupling of two and three QRKRs.

Unfortunately, exact numerical study of the many-body QRKR model is not feasible presently. However, at high momenta, it can be approximated by the integrable many-body LQKR model, and this approximation works only better as the system goes to higher momentum states. In Ref.~[\onlinecite{Keser16}], the many-body LQKR model was analytically shown to exhibit the DMBL phase. Besides that, as opposed to the case of QKR, the classical model behind QRKR is not chaotic at high momenta. As we have shown, in the cases of two and three coupled QRKRs, localization has a quantum origin and does not rely on the existence of KAM tori in the phase space of the corresponding classical problem. However, for a large number of interacting rotors, if this localization happens to deteriorate completely, and the growth of the particles' momenta recovers, at high enough momenta the system will enter the integrable regime and get localized. This is our main argument in favor of DMBL in the nonintegrable system of the many-body QRKR model. In general, the observable dynamical localization can represent a nontrivial interplay of both effects that may be difficult to disentangle. In summary, our argument supplemented by few-body calculations provide a strong hint that the nonintegrable many-body QRKR model should exhibit dynamically localized many-body states.

In addition, we propose a class of kicked lattice models that map onto various kicked-rotor models and can be realized in the framework of cold atoms in optical lattices. This realization might allow one to study dynamical localization including DMBL, and other anomalous transport phenomena exhibited by the QRKR and its many-body versions in experiment.

\begin{acknowledgments}
This work was supported by US-ARO (Contract No.~W911NF1310172), NSF-DMR 1613029, and the Simons Foundation. The authors are grateful to Aydin Cem Keser, Gil Refael, Trey Porto, Steve Rolston, Varun Vaidya, Borzumehr Toloui, and Victor Yakovenko for valuable discussions.
\end{acknowledgments}

\end{document}